\def\year{2016}\relax
\documentclass[letterpaper]{article}
\usepackage{aaai16}
\usepackage{times}
\usepackage{helvet}
\usepackage{courier}
\usepackage{amsfonts}
\usepackage[latin1,utf8]{inputenc}
\usepackage{mathrsfs}
\usepackage{amsmath}

\newtheorem{thm}{Theorem}
\newtheorem{defn}[thm]{Definition}
\newtheorem{cor}[thm]{Corollary}
\newtheorem{pro}[thm]{Proposition}
\newtheorem{lem}[thm]{Lemma}
\newtheorem{obs}[thm]{Observation}
\newtheorem{exa}[thm]{Example}
\newenvironment{theorem}      {\begin{thm} \em}{\end{thm}}
\newenvironment{definition}   {\begin{defn} \em}{\end{defn}}

\newenvironment{lemma}        {\begin{lem} \em}{\end{lem}}
\newenvironment{observation}  {\begin{obs} \em}{\end{obs}}

\newcommand{\LP}{\cal L}

\newcommand{\scont}{-}

\newcommand{\fcont}[2]{{#1} {\scont} { #2}}

\newcommand{\mypostnoname}[2]{\noindent {\bf (#1)}\hspace{2mm}{#2}\hfill{\hspace*{0,1cm}}\\ }
\newcommand{\qed}{\hfill \rule{2mm}{2mm}}
\nocopyright
\begin{document}

\title{Studies on Brutal Contraction and Severe Withdrawal: Preliminary Report}
\author{Marco Garapa \\Universidade da Madeira\\CIMA - Centro de Investiga\c c\~ao\\ em Matem\'atica e Aplica\c c\~oes\thanks{Supported by FCT - Funda\c c\~ao para a Ci\^encia e a Tecnologia through project UID/MAT/04674/2013 (CIMA).}\\marco@uma.pt \And Eduardo Ferm\'e \\Universidade da Madeira\\NOVA Laboratory for Computer Science \\and Informatics (NOVA LINCS)\thanks{Supported by FCT MCTES and NOVA LINCS UID/CEC/04516/2013.}\\ferme@uma.pt\And Maur\'{i}cio D. L. Reis\\Universidade da Madeira\\CIMA - Centro de Investiga\c c\~ao\\ em Matem\'atica e Aplica\c c\~oes$^*$\\m\_reis@uma.pt}
\maketitle

\begin{abstract}
In this paper we study the class of brutal base contractions that are based on a bounded ensconcement and also the class of severe withdrawals which are based on bounded epistemic entrenchment relations that are defined by means of bounded ensconcements (using the procedure proposed by Mary-Anne Williams). We present axiomatic characterizations for each one of those classes of functions and investigate the interrelation among them.
\end{abstract}

\section{Introduction}
The central goal underlying the research area of {\em logic of theory change} is the study of the changes which can occur in the belief state of a rational agent when he receives new information.

The most well known model of theory change was proposed by \citeauthor{AGM85} \shortcite{AGM85} and is, nowadays, known as the AGM model. Assuming that the belief state of an agent is modelled by a {\em belief set} (i.e. a logically closed set of sentences), this framework essentially provides a definition for contractions --- i.e. functions that receive a sentence (representing the new information received by the agent), and return a belief set which is a subset of the original one that does not contain the received sentence. In the mentioned paper, the class of {\em partial meet contractions} was introduced and axiomatically characterized. Subsequently several constructive models have been presented for the class of contraction functions proposed in the AGM framework (such as the {\em system of spheres-based contractions} \cite{Gro88}, {\em safe/kernel contractions} \cite{AM85,Han94}, and the {\em epistemic entrenchment-based contractions} \cite{Gar88,GM88}). Also several adaptations and variations of those constructive models have been presented and studied in the literature as it is the case, for example, of {\em severe withdrawals} (or mild contractions or Rott's contractions) \cite{Rot91,RP99} which results of {\em simplifying} the definition of {\em epistemic entrenchment-based contractions}.

Although the AGM model has quickly acquired the status of standard model of theory change, several researchers (for an overview see \cite{FH11}) have pointed out its inadequateness in several contexts and proposed several extensions and generalizations to that framework. One of the most relevant of the proposed extensions of the AGM model of contraction is to use sets of sentences not (necessarily) closed under logical consequence --- which are designated belief bases --- rather than belief sets to represent belief states.

Hence, several of the existing models (of AGM contractions) were generalized to the case when belief states are represented by belief bases instead of belief sets. Among those we emphasize the {\em ensconcement-based contractions} and the {\em brutal base contractions} (of belief bases) proposed in \cite{Wil95}, which can be seen as adaptations to the case of belief bases of the {\em epistemic entrenchment-based contractions} and of the {\em severe withdrawals}, respectively.
In fact, the definitions of those operations are both based on the concept of {\em ensconcement}, which is an adaptation of the concept of epistemic entrenchment relation to the case of belief bases. In the mentioned paper Mary-Anne Williams has also presented a method for constructing an epistemic entrenchment from an ensconcement relation.

In the present paper we will study the interrelation among {\em brutal base contractions} (of belief bases) and {\em severe withdrawals} (of belief sets). More precisely, we will devote special attention to the class of {\em brutal base contractions} which are based on bounded ensconcements --- the so-called {\em bounded brutal base contractions} --- and also to the class of the so-called {\em ensconcement-based severe withdrawals}, which is formed by the severe withdrawals that are based on an epistemic entrenchment relation defined from a bounded ensconcement using Mary-Anne William's method. We shall provide axiomatic characterizations to each one of those classes of functions and study the interrelation among them.

This paper is organized as follows: Firstly we provide the notation and background needed for the rest of the paper. After that we provide axiomatic characterizations for the classes of {\em bounded brutal base contractions} and of {\em ensconcement-based severe withdrawals}. Furthermore we show how to define a {\em bounded brutal base contraction} from an {\em ensconcement-based severe withdrawal} and vice-versa. Finally,  we briefly summarize the main contributions of the paper. In the appendix we provide proofs for the theorems. Proofs for all the remaining results are available at http://www.cee.uma.pt/ferme/GFR16-full.pdf.

\section{Background}\label{background}

\subsection{Formal preliminaries}
We will assume a language $\LP$ that is closed under truth-functional operations and a consequence operator $Cn$ for $\LP$. $Cn$ satisfies the standard Tarskian properties, namely {inclusion} ($A \subseteq Cn(A)$), {monotony} (if $A \subseteq B$, then  $Cn(A) \subseteq Cn(B)$), and {iteration} ($Cn(A) = Cn(Cn(A))$).  It is supraclassical and compact, and satisfies {deduction} (if $\beta \in Cn(A \cup \{ \alpha \})$, then $(\alpha \rightarrow \beta) \in Cn(A)$).  $A \vdash \alpha$ will be used as an alternative notation for $\alpha \in Cn(A)$, $\vdash \alpha$ for $\alpha \in Cn(\emptyset)$ and $Cn(\alpha)$ for $Cn(\{\alpha\})$.  Upper-case letters denote subsets of $\LP$. Lower-case Greek letters denote elements of $\LP$. \\
A well-ranked preorder on a set $X$ is a preorder such that every nonempty subset of $X$ has a minimal member, and similarly an inversely well-ranked preorder on a set $X$ is a preorder such that every nonempty subset of $X$ has a maximal member. A total preorder on $X$ is bounded if and only if it is both well-ranked and inversely well-ranked.\footnote{In \cite{Wil94b} a preorder in these conditions is designated by {\em finite}, however we think it is more adequate to use the denomination {\em bounded}.}

\subsection{AGM}
The AGM model of belief change was proposed by \citeauthor{AGM85} \shortcite{AGM85} and acquired the status of standard model of belief change. In this model beliefs are represented by a set of sentences closed under logical consequence. In the AGM framework there are three operations to be considered, namely expansion, contraction and revision. Expansion, consists of adding new information (represented by sentences) in the original set preserving logical closure. Contraction, consists of eliminating sentences from a belief set, in such a way that the remaining set does not imply a specified sentence. Revision, consists in incorporating a sentence in the original set, but (eventually) eliminating some sentences in order to retain consistency of the revised set. AGM has been characterized in, at least five, different ways: {\em
Postulates}, {\em partial meet functions}, {\em epistemic entrenchment},{\em safe/kernel contraction} and {\em Grove' sphere-systems} (for an overview see \cite{FH11}).\\
One of the Postulates included in the axiomatic characterization of the contraction operator is recovery:

 \

 \mypostnoname{Recovery}   {$K\subseteq(K-\alpha)+\alpha$} {}

{\em Recovery} is based in the principle that ``it is reasonable to require that we get all of the beliefs [...] back again after first contracting and then expanding with respect to the same belief" \cite{Gar82}. Nevertheless, the {\em recovery} postulate have been criticized by several authors \cite{Fuh91,Han91b,Lev91,Nie91} as a general principle that contractions should hold.  Alternative contraction models were proposed in which the {\em recovery} postulate does not hold, for instance: Levi Contraction \cite{Lev91}, Severe Withdrawal \cite{Rot91,RP99} and Semi-contraction \cite{Fer98}.

\subsection{Epistemic Entrenchment}

Epistemic entrenchment was introduced in \cite{Gar88,GM88} and relies on the idea that contractions on a belief set $K$ should be based on an ordering of its sentences according to their epistemic entrenchment. When a belief set $K$ is contracted it is prefered to give up beliefs with lower entrechment over others with a higher entrechment.  G\"{a}rdenfors proposed the following set of axioms that an epistemic entrechment order $\leq$ related to a belief set $K$ should satisfy:\\

\mypostnoname{EE1} {If $\alpha \leq \beta$ and $\beta \leq \gamma$, then $\alpha \leq \gamma$ (Transitivity)} {}
\mypostnoname{EE2} {If $\alpha \vdash \beta$, then $\alpha \leq \beta$ (Dominance)}  {}
\mypostnoname{EE3} {$\alpha \leq (\alpha \wedge \beta)$ or $\beta \leq (\alpha \wedge \beta)$ (Conjunctiveness)} {} {\bf }
\mypostnoname{EE4} {If $K\not \vdash \perp$, then $\alpha \not \in K$ if and only if $\alpha \leq \beta$ for all $\beta$ (Minimality)} {}
\mypostnoname{EE5} {If $\beta \leq \alpha$ for all $\beta$, then $\vdash \alpha$ (Maximality)} {}

If $\leq$ is well-ranked and inversely well-ranked, then the epistemic entrenchment is well-ranked and inversely well-ranked, and therefore is a bounded epistemic entrenchment.
The relation $\leq$ of epistemic entrenchment is independent of the change functions in the sense that it does not refer to any contraction or revision function. In addition to stating the axioms of entrenchment,  G\"ardenfors  proposed the following entrenchment-based contraction functions:\\

\mypostnoname{$G_\leq$} { $\beta \in \fcont{K}{\alpha}$ if and only if ${\beta} \in {K}$ and, either $\vdash {\alpha}$ or ${\alpha} < ({\alpha} \vee  {\beta})$} {}

The crucial clause of $(G_\leq)$ is  ${\alpha} < ({\alpha} \vee  {\beta})$. This clause can be justified with reference to the recovery postulate \cite{GM88}.

\subsubsection{Severe withdrawal:}

\citeauthor{Rot91} \shortcite{Rot91} proposed a more intuitive alternative definition, later called {\em Severe withdrawal} (or mild contraction or Rott's contraction) \cite{RP99}:\\

\mypostnoname{$R_\leq$} { $\beta \in \fcont{K}{\alpha}$ if and only if ${\beta} \in {K}$ and, either $\vdash {\alpha}$ or ${\alpha} < {\beta}$} {}

\citeauthor{CL06} \shortcite{CL06} have analyzed it in terms of minimal loss of informational value.
It has been shown to satisfy the implausible postulate of expulsiveness. (If $\not \vdash \alpha$ and $\not \vdash \beta$, then either $\alpha \not \in K \div \beta$ or  $\beta \not \in K \div \alpha$) \cite{Han99b}. \citeauthor{LR91} \shortcite{LR91} abstained from recommending either a particularly expulsive contraction (severe withdrawal) or a particularly retentive one (AGM contraction). They argued that these extremes should be taken as ``upper" and ``lower" bounds and that any ``reasonable" contraction function should be situated between them. This condition was called the Lindstr{\"o}m's  and Rabinowicz's interpolation thesis \cite{Rot95}. Severe  withdrawal was axiomatized independently by \citeauthor{RP99} \shortcite{RP99} and by \citeauthor{FR98} \shortcite{FR98}. The following set of postulates  characterize severe withdrawals \cite{RP99}:\\

\mypostnoname{$\div 1$} { $K\div\alpha=Cn(K\div\alpha)$}{}
\mypostnoname{$\div 2$} { $K\div\alpha\subseteq K$} {}
\mypostnoname{$\div 3$} { If $\alpha \not \in K$ or $\vdash \alpha$, then $ K\subseteq K \div\alpha$} {}
\mypostnoname{$\div 4$} { If $\not\vdash \alpha$, then $\alpha \not \in K\div \alpha$} {}
\mypostnoname{$\div 6$} { If $Cn(\alpha)=Cn(\beta)$, then $K\div \alpha=K\div \beta$} {}
\mypostnoname{$\div 7a$} { If $\not\vdash \alpha$, then $K\div \alpha \subseteq K\div (\alpha \wedge \beta)$} {}
\mypostnoname{$\div 8$} { If $\alpha \not \in K\div (\alpha \wedge \beta)$, then $K\div (\alpha \wedge \beta) \subseteq K \div \alpha$} {}

Severe withdrawal also satisfies the following postulates:\\

\mypostnoname{$\div 10$}{ If $\not \vdash \alpha$ and $\alpha \in K\div\beta$, then $K\div\alpha\subseteq K\div\beta$.}{}
\mypostnoname{Linearity}{Either $K\div \alpha \subseteq K\div \beta$ or $K\div \beta \subseteq K\div \alpha$.}{}
\mypostnoname{Expulsiveness}{If $\not \vdash \alpha$ and $\not \vdash \beta$, then either $\alpha \not \in K \div \beta$ or  $\beta \not \in K \div \alpha$.}{}

\citeauthor{RP99} \shortcite{RP99} showed that an alternative axiomatization of severe withdrawals consists of the postulates $(\div 1)$ to $(\div 4)$ and $(\div 6)$ and:\\

\mypostnoname{$\div 9$}{If $\alpha \not \in K\div \beta$, then $K\div \beta \subseteq K\div \alpha$.}{}

\subsection{Ensconcement}

\citeauthor{Wil95} \shortcite{Wil92,Wil95} defines an {\em ensconcement} relation on a belief base A as a transitive and connected relation $\preceq$ that satisfies the following three conditions:\footnote{ $\alpha \prec \beta$ means  $\alpha \preceq \beta$ and $\beta \not \preceq \alpha$. $\alpha =_\preceq \beta$ means  $\alpha \preceq \beta$ and $\beta \preceq \alpha$.}\\

\mypostnoname{$\preceq$1}{ If $\beta \in A\setminus Cn(\emptyset)$, then $\{\alpha \in A: \beta \prec \alpha \} \not \vdash \beta$}{}
\mypostnoname{$\preceq$2}{If $\not \vdash \alpha$ and $\vdash \beta$, then $\alpha \prec \beta$, for all $\alpha,\beta\in A$}{}
\mypostnoname{$\preceq$3}{If $\vdash \alpha$ and $\vdash \beta$, then $\alpha \preceq \beta$, for all $\alpha,\beta\in A$}{}

$(\preceq 1)$ says that the formulae that are strictly more ensconced than $\alpha$ do not (even conjointly) imply $\alpha$. Conditions $(\preceq 2)$ and $(\preceq 3)$ say that tautologies are the most ensconced formulae. If $\preceq$ is well-ranked/inversely well-ranked, then the ensconcement $(A,\preceq)$ is well-ranked/inversely well-ranked. If $\preceq$ is both well-ranked and inversely well-ranked then it is a bounded ensconcement.

\

Given an ensconcement relation, a cut operator for $\alpha \in Cn(A)$ is defined by:

\

$cut _{\preceq}(\alpha)=\{\beta \in A: \{\gamma \in A:\beta \prec \gamma \} \not \vdash \alpha\}$.

\

A proper cut for $\alpha \in \LP$ is defined by:

\

$cut_{\prec}(\alpha) = \{\beta \in A: \{ \gamma \in A: \beta \preceq
\gamma\} \not \vdash \alpha \}$

\

\begin{observation} \cite{Wil94b}\label{cut_with_alpha}

If $\alpha \in A$, \ $cut_{\prec}(\alpha) = \{\beta \in A: \alpha \prec
\beta\}  $

\end{observation}

The previous observation says that when $\alpha$ is an explicit belief, its proper cut is the subset of $A$ such that its members are strictly more ensconced than $\alpha$. Other properties of {\em proper cut} are:

\begin{observation}\label{preTeoA}
Let $(A,\preceq)$ be a bounded ensconcement and $\alpha,\beta\in Cn(A)$, then:
\begin{description}
\item [(a)] Let $\not \vdash \beta$. If $cut_{\prec}(\alpha)\subseteq cut_{\prec}(\beta)$, then $cut_{\preceq}(\alpha)\subseteq cut_{\preceq}(\beta)$.
\item [(b)] If $\vdash \beta$ and $\not \vdash \alpha$, then $cut_{\preceq}(\beta)\subset cut_{\preceq}(\alpha)$.
\end{description}
\end{observation}

Intuitively, an ensconcement is to belief bases as epistemic entrenchment is to belief sets. Williams explores this relation:

\begin{definition}\cite{Wil94} \label{MAWdef2}
Let $(A,\preceq)$ be an ensconcement. For $\alpha,\beta \in L$, define $\leq_{\preceq}$ to be given by: $\alpha \leq_{\preceq} \beta$ if and only if either:\\
i) $\alpha \not \in Cn(A)$, or
ii) $\alpha, \beta \in Cn(A)$ and $cut_{\preceq}(\beta)\subseteq cut_{\preceq}(\alpha)$.
\end{definition}

\begin{observation}\cite{Wil94}\label{Teo4MAW}
If $(A,\preceq)$ is an ensconcement, then $\leq_{\preceq}$ is an epistemic entrenchment related to $Cn(A)$.
\end{observation}

\begin{observation}\cite{Wil94} \label{CorolMAW}
Given an ensconcement $(A,\preceq)$, $\preceq$ is well-ranked (inversely well-ranked, bounded) if and only if $\leq_{\preceq}$ is well-ranked (inversely well-ranked, bounded).
\end{observation}

\subsection{Brutal Contraction}

Mary-Anne Williams \cite{Wil94} defines two operators for base contraction: The first one inspired in AGM contraction (ensconcement-based contraction) and the second one inspired in severe withdraw (brutal contraction). In this paper we will focus in the second one. Brutal contraction, as Mary-Anne Williams says, ``retains as little as necessary of the theory base".

\begin{definition}\cite{Wil94} \label{brutal1}
Let A be a belief base. An operation $-$ is a brutal base contraction on $A$ if and only if there is an ensconcement relation $\preceq$ on $A$ such that:

\

$\beta \in  A - \alpha$ if and only if $\beta \in A$ and either (i)
 $\alpha \in Cn(\emptyset)$ or (ii) $\beta \in cut_{\prec}(\alpha)$

\end{definition}

In \cite{GFR16b} the following axiomatic characterization for brutal base contractions was presented:

\begin{observation}\cite{GFR16b} \label{rep1}
Let $A$ be a belief base. An operator $-$ of $A$ is a brutal base contraction on $A$ if and only if it satisfies:\\

\mypostnoname{Success} {If $\not \vdash \alpha$, then $A-\alpha \not \vdash \alpha$}{}
\mypostnoname{Inclusion} {$ A - \alpha \subseteq A$}{}
\mypostnoname{Vacuity} {If $A \not \vdash \alpha$, then  $ A  \subseteq A-\alpha$}{}
\mypostnoname{Failure} {If $\vdash \alpha$, then $A - \alpha = A$}{}
\mypostnoname{Relative Closure} {$A \cap Cn(A-\alpha)\subseteq A-\alpha$}{}
\mypostnoname{Strong Inclusion}{If $A-\beta \not \vdash \alpha$, then $A-\beta\subseteq A-\alpha$}
\mypostnoname{Uniform Behaviour}{If $\beta \in A$, $A \vdash \alpha $ and $A-\alpha=A-\beta$, then $\alpha \in Cn(A- \beta \cup \{\gamma \in A: A- \beta = A-\gamma\})$}{}
\end{observation}

The following observation lists some other well-known postulates which are satisfied by the brutal base contraction functions.

\begin{observation}\cite{GFR16b} \label{afterpost} Let $A$ be a belief base and $-$ an operator on $A$ that satisfies {\em success, inclusion, vacuity, failure, relative closure, strong inclusion} and {\em uniform behaviour}. Then $-$ satisfies:
\begin{description}
\item[(a)] If $\alpha \in A \setminus A-\beta$, then $A-\beta\subseteq A-\alpha$.
\item[(b)] If $A-\alpha \subset A- \beta$, then $A-\beta \vdash \alpha$.
\item[(c)] If $\vdash \alpha$ and $\alpha \in A$, then $\alpha \in A- \beta$.
\item[(d)] If $\vdash \alpha \leftrightarrow \beta$, then $ A - \alpha = A - \beta$. (Extensionality)
\end{description}
\end{observation}

\section{Bounded Brutal Base Contraction Functions}

In this subsection we introduce the bounded brutal base contractions and obtain an axiomatic characterization for that class of functions.

\begin{definition}\label{brutal2} Let $A$ be a belief base. An operation $-$ is a bounded brutal base contraction on $A$ if and only if it is a brutal base contraction based on a bounded ensconcement.
\end{definition}

We introduce the following postulates:\\

\mypostnoname{Upper Bound} {For every non-empty set $X\subseteq A$ of nontautological formulae, there exists $\alpha \in X$ such that $A-\beta \subseteq A-\alpha$ for all $\beta \in X$}{}
\mypostnoname{Lower Bound} {For every non-empty set $X\subseteq A$ of nontautological formulae, there exists $\alpha \in X$ such that $A-\alpha \subseteq A-\beta$ for all $\beta \in X$}{}
\mypostnoname{Clustering} {If $\beta \in A$, then there exists $\alpha \in A \cup Cn(\emptyset)$ such that $A-\alpha=A-\beta \cup \{\gamma \in A: A-\beta=A-\gamma\}$}{}

{Upper Bound} (respectively { Lower Bound}) states that every non-empty set of nontautological formulae of $A$ contains an element which is such that the result of contracting $A$ by that sentence is a superset (respectively a subset) of any set which results of contracting $A$ by one of the remaining sentences of that subset.

{ Clustering} asserts that for any sentence $\beta$ in $A$ there exists some sentence $\alpha$ in $A \cup Cn(\emptyset)$ such that the result of the contraction of $\alpha$ from $A$ is the set consisting of the union of the result of contracting $A$ by $\beta$ with the set formed by all the sentences of $A$ which are such that the result of contracting it from $A$ coincides with the result of contracting $A$ by $\beta$.

The two following observations present some interrelations among the above proposed postulates and some of the of the postulates included in the axiomatic characterization that was obtained for the class of brutal base contraction.

\begin{observation} \label{s2lemma1}
Let $A$ be a belief base and $-$ an operator on A that satisfies {\em success, inclusion, failure, relative closure, strong inclusion} and {\em lower bound}. Then $-$ satisfies {\em clustering}.
\end{observation}

\begin{observation} \label{s2lemma2}
Let $A$ be a belief base and $-$ an operator on A that satisfies {\em failure, success, strong inclusion and clustering}. Then $-$ satisfies {\em uniform behaviour}.
\end{observation}

We are now in a position to present an axiomatic characterization for the class of bounded brutal base contractions.

\begin{theorem} \label{AxchFBC}  (Axiomatic characterization of bounded brutal base contraction functions)
Let $A$ be a belief base. An operator $-$ on $A$ is a bounded brutal base contraction on $A$ if and only if it satisfies {\em success, inclusion, vacuity, failure, relative closure, lower bound, upper bound} and {\em strong inclusion}.
\end{theorem}

The following observation exposes another relevant property of the bounded brutal base contractions which will be useful further ahead. More precisely, it asserts that for any non-tautological sentence $\alpha$ which is deducible from $A$ it holds that the result of contracting $A$ by $\alpha$ coincides with the result of the contraction of $A$ by some sentence explicitly included in $A$.

\begin{observation} \label{lemma IMP}
Let $A$ be a belief base and $-$ an operator on $A$ that satisfies \em{success, inclusion, failure, relative closure, strong inclusion} and \em{lower bound}. Then $-$ satisfies:\\
For all $\alpha \in Cn(A)\setminus Cn(\emptyset)$ there exists $\beta \in A$ such that $A-\alpha=A-\beta$.
\end{observation}

\section{Relation between Bounded Brutal Base Contraction and Ensconcement-based Severe Withdrawal}

In this section we will define and axiomatically characterize a particular kind of severe withdrawals which we will show to be the contraction functions that correspond to the bounded brutal base contractions in the context of belief set contractions.

We start by noticing that, given a bounded ensconcement $(A,\preceq)$, we can combine Definitions $\ref{MAWdef2}$ and ($R_\leq$) in order to define a contraction function on the belief set $Cn(A)$. This kind of functions is formally introduced in the following definition.

\begin{definition}
$\div$ is an ensconcement-based withdrawal related to $(A,\preceq)$ if and only if $(A,\preceq)$ is a bounded ensconcement such that $Cn(A)\div \alpha=Cn(A)\div_{\leq_{\preceq}} \alpha$, where $\leq_{\preceq}$ is the epistemic entrenchment with respect to $Cn(A)$ defined by Definition \ref{MAWdef2} and $\div_{\leq_{\preceq}}$ is the severe withdrawal on $Cn(A)$ defined by ($R_\leq$).
\end{definition}

Comparing the above definition with Definitions $\ref{brutal1}$ and $\ref{brutal2}$ it becomes clear that there is a strong interrelation among the ensconcement-based severe withdrawals and the (bounded) brutal base contractions. That interrelation is explicitly presented in the two following theorems. More precisely, given a bounded ensconcement $(A,\preceq)$, these two results expose how the $\preceq$-based brutal contraction on $A$ can be defined from the ensconcement-based withdrawal related to $(A,\preceq)$ and, vice-versa, how the latter can be defined by means of the former.

\begin{theorem} \label{theo_B}
Let $(A,\preceq)$ be a bounded ensconcement, $-$ be the $\preceq$-based brutal contraction, and $\div_{\leq_{\preceq}}$ be the ensconcement-based severe withdrawal related to $(A,\preceq)$, then $A-\alpha=(Cn(A)\div_{\leq_{\preceq}}\alpha)\cap A$.
\end{theorem}

\begin{theorem} \label{theo_A}
Let $(A,\preceq)$ be a bounded ensconcement, $-$ be the $\preceq$-based brutal contraction, and $\div_{\leq_{\preceq}}$ be the ensconcement-based severe withdrawal related to $(A,\preceq)$, then $Cn(A)\div_{\leq_{\preceq}}\alpha=Cn(A-\alpha)$.
\end{theorem}

\subsection{Axiomatic Characterization of ensconcement-based severe withdrawals}

\

In this subsection we will present an axiomatic characterization for the class of ensconcement-based severe withdrawals. To do that we must start by introducing the following postulate:

\mypostnoname{Base-reduction} {If $Cn(A)\div \alpha \vdash \beta$, then $(Cn(A)\div\alpha)\cap A\vdash \beta$}{}

This postulate essentially states that that the result of contracting the belief set $Cn(A)$ by any sentence $\alpha$ coincides with the logical closure of some subset of $A$.
Indeed, it is not hard to see that base-reduction is equivalent to the following postulate: $\forall \alpha \exists A'\subseteq A: Cn(A')=Cn(A)\div \alpha$ (which is very similar to the postulate of {\em finitude} proposed by \citeauthor{Han98} \shortcite{Han98}).

The following observation highlights that for a severe withdrawal that satisfies the postulates of {\em base-reduction} and {\em lower bound} it also holds that for any non-tautological sentence $\alpha$ in $Cn(A)$ the result of the contraction of $Cn(A)$ by $\alpha$ coincides with the result of the contraction of $Cn(A)$ by some sentence in $A$.

\begin{observation} \label{AAZ}
Let $\div$ be an operator on $Cn(A)$ that satisfies  $(\div 1)$, $(\div 2)$, $(\div 4)$, ($\div 9$), {\em base-reduction} and {\em lower bound}, then for all $\alpha \in Cn(A)\setminus Cn(\emptyset)$ there exists $\beta\in A$ such that $Cn(A)\div \alpha=Cn(A)\div \beta$.
\end{observation}

We are now in a position to present the following axiomatic characterization for the ensconcement-based severe withdrawals.

\begin{theorem} \label{theo_C}
Let $A$ be a belief base and $\div$ be an operator on $Cn(A)$. $\div$ satisfies  $(\div 1)$ to $(\div 4)$, $(\div 6)$, ($\div 9$), {\em base-reduction, upper bound} and {\em lower bound} if and only if there exists a bounded ensconcement such that $\div$ is an ensconcement-based withdrawal related to $(A,\preceq)$.
\end{theorem}

Theorems $\ref{theo_B}$ and $\ref{theo_A}$ expose how a base contraction function can be defined from a belief set contraction function and, vice-versa. Combining those two results with the axiomatic characterizations presented in Theorems $\ref{AxchFBC}$ and $\ref{theo_C}$ we can obtain the following results which highlight the correspondence among sets of postulates for base contraction and sets of postulates for belief set contraction.

\begin{cor} \label{theo_E}
An operator $-$ on $A$ satisfies {\em success, inclusion, vacuity, failure, relative closure, strong inclusion, upper bound} and {\em lower bound} if and only if there exists an operator $\div$ on $Cn(A)$ that satisfies  $(\div 1)$ to $(\div 4)$, $(\div 6)$, ($\div 9$), {\em base-reduction, upper bound} and {\em lower bound} such that: $A-\alpha=Cn(A\div \alpha)\cap A$.
\end{cor}

\begin{cor} \label{ultimo}
An operator $\div$ on $Cn(A)$ satisfies $(\div 1)$ to $(\div 4)$, $(\div 6)$, ($\div 9$), {\em base-reduction, upper bound} and {\em lower bound} if and only if there exists an operator $-$ on $A$ that satisfies {\em success, inclusion, vacuity, failure, relative closure, strong inclusion, upper bound} and {\em lower bound} such that: $Cn(A)\div \alpha=Cn(A-\alpha)$.
\end{cor}

The two following observations consist of a slight refinement of the right to left part of Corollary $\ref{ultimo}$. More precisely these results specify more precisely which properties of the belief base contraction are needed in order to assure that the belief set contraction obtained from it as exposed in Theorem $\ref{theo_A}$ satisfies certain postulates.

\begin{observation} \label{ultimo2}
Let $A$ be a belief base and $-$ be an operator on $A$ that satisfies {\em success, inclusion, vacuity, failure, relative closure} and {\em strong inclusion}. If $\div$ is an operator on $Cn(A)$ defined by $Cn(A)\div \alpha=Cn(A-\alpha)$ then $\div$ satisfies  $(\div 1)$ to $(\div 4)$, $(\div 6)$, ($\div 9$) and {\em base-reduction}.
\end{observation}

\begin{observation} \label{ultimo3}
Let $A$ be a belief base and $-$ be an operator on $A$ that satisfies {\em success, inclusion, failure, relative closure, upper bound, lower bound} and {\em strong inclusion}. If $\div$ is an operator on $Cn(A)$ defined by $Cn(A)\div \alpha=Cn(A-\alpha)$ then $\div$ satisfies {\em upper bound} and {\em lower bound}.
\end{observation}

\section{Conclusions}
We have presented an axiomatic characterizations for the subclass of brutal base contractions formed by the brutal contractions that are based on a bounded ensconcement relation. We have also introduced and axiomatically characterized the class of ensconcement-based severe withdrawals which is formed by the severe withdrawals that are based on epistemic entrenchment relations which are obtained from an ensconcement relation using the construction proposed by Mary-Anne Williams. Some results were presented concerning the interrelation among the classes of bounded brutal base contractions and of ensconcement-based severe withdrawals. Finally we presented some results relating base contraction postulates and belief set contraction postulates by means of explicit definitions of belief set contractions from base contractions and vice-versa.

\section*{Acknowledgements}

We wish to thank the three reviewers for their comments which have contributed to the improvement of this paper.

\section*{Appendix: Proofs}

\subsection*{Previous Lemmas}
\begin{lemma}  \cite{FKR08}\label{lemprev1}
\begin{description}
\item[(a)] If $\not \vdash \alpha, cut_{\prec}(\alpha) \not \vdash \alpha$.
\item[(b)] If $A \not \vdash \alpha, cut_{\prec}(\alpha) = A$.
\item[(c)] If $\beta \vdash \alpha$, then $cut_{\prec}(\alpha) \subseteq cut_{\prec}(\beta)$.
\item[(d)] If $\alpha \preceq \beta$, then $cut_{\prec}(\beta) \subseteq cut_{\prec}(\alpha)$.
\item[(e)] If $cut_{\prec}(\alpha) \vdash \beta$, then $cut_{\prec}(\alpha \wedge \beta) = cut_{\prec}(\alpha)$.
\item[(f)] If $cut_{\prec}(\alpha) \not \vdash \beta$, then $cut_{\prec}(\alpha \wedge \beta) = cut_{\prec}(\beta)$.

\end{description}
\end{lemma}

\begin{lemma} \cite[Observation 19(ii)]{RP99} \label{obser.19ii}
If $\div$ is a severe withdrawal function, then $\div$ can be represented as an entrenchement-based withdrawal where the relation $\leq$ on which $\div$ is based is obtained by\\
(Def $\leq$ from $\div$) $\alpha \leq \beta$ if and only if $\alpha \not \in K\div \beta$ or $\vdash \beta$\\
and $\leq$ satisfies (EE1) to (EE5).
\end{lemma}

\begin{lemma} \label{le1_theo3}
Let $(A,\preceq)$ be a bounded ensconcement and $cut_{\preceq}(\alpha)\not = \emptyset$. Then there exists $\beta \in cut_{\preceq}(\alpha)$ such that $cut_{\preceq}(\beta)=cut_{\preceq}(\alpha)$.
\end{lemma}

\begin{lemma} \label{le2_theo3}
Let $(A,\preceq)$ be a bounded ensconcement and $\alpha \in Cn(A)$. Then $cut_{\preceq}(\alpha) \vdash \alpha$.
\end{lemma}

\begin{lemma} \label{obs1}
Let $(A,\preceq)$ be a bounded ensconcement and $\alpha,\beta\in Cn(A)$. If $cut_{\prec}(\alpha) \subset cut_{\prec}(\beta)$, then $cut_{\preceq}(\alpha) \subset cut_{\preceq}(\beta)$.
\end{lemma}

\subsection*{Proofs}

{\bf Proof of Theorem \ref{AxchFBC}}\\
From bounded brutal base contraction to postulates\\
Let $-$ be a bounded brutal base contraction operator on $A$. By Observation \ref{rep1} $-$ satisfies {\em success, inclusion, vacuity, failure, relative closure} and {\em strong inclusion}. It remains to show that $-$ satisfies {\em upper bound} and {\em lower bound}.\\
{\bf Upper Bound} Let $X\subseteq A$ be a non empty set of non-tautological formulae. Since $\preceq$ is well ranked there exists $\beta \in X$ such that $\beta \preceq \alpha$ for all $\alpha \in X$. Hence, by Lemma \ref{lemprev1} (d), there exists $\beta \in X$ for all $\alpha \in X$ such that $cut_{\prec}(\alpha)\subseteq cut_{\prec}(\beta)$. Therefore, by definition of $-$ there exists $\beta \in X$ for all $\alpha \in X$ such that $A-\alpha\subseteq A-\beta$. \\
{\bf Lower Bound} Analogous to {\em upper bound}.\\
From postulates to bounded brutal base contraction\\
Let $-$ be an operator on $A$ that satisfies {\em success, inclusion, vacuity, failure, relative closure, lower bound, upper bound} and {\em strong inclusion}. From Observation \ref{s2lemma1} and Observation \ref{s2lemma2} it follows that $-$ satisfies {\em uniform behaviour}. Let $\preceq$ be defined by:\\
$\alpha \preceq \beta $ iff $ \left\{ \begin{array}{l}
     A-\beta \subseteq A- \alpha$ and $ \not \vdash \alpha \\ $ or$\\
    \vdash \beta \\
\end{array}\right.$ \\ \\

According to the Postulates to Construction part of the proof of Observation \ref{rep1} $\preceq$ satisfies {\bf $(\preceq 1)$ - $(\preceq 3)$} and is such that \\

$A - \alpha = \left\{
\begin{array}{ll}
 \ cut_{\prec}(\alpha)  & $if $  \not \vdash \alpha \\
A& $otherwise$\\
 \end{array}
\right. $ \\
It remains to prove that $\preceq$ is bounded. To do so we must prove that $\preceq$ is well-ranked and inversely well-ranked.\\
{\bf ($\preceq$ is well-ranked)} Let $X\not =\emptyset$ and $X\subseteq A$. We will prove by cases:\\
Case 1) All formulae in $X$ are tautologies. Let $\beta$ be one of those formulas. Hence by ($\preceq$ 3) $\beta \preceq \alpha$ for all $\alpha \in X$.\\
Case 2) All formulae in $X$ are non-tautological. By {\em upper bound} there exists $\beta \in X$ such that $A-\alpha \subseteq A-\beta$ for all $\alpha \in X$. Hence, by definition of $\preceq$, there exists $\beta \in X$ such that $\beta \preceq \alpha$ for all $\alpha \in X$.\\
Case 3) There are some formulae in $X$, that are tautological and others that are not. Consider $X'=X\setminus Cn(\emptyset)$. Hence, by the previous case, there exists $\beta \in X'$ such that $\beta \preceq \alpha'$ for all $\alpha' \in X'$. Therefore, it follows from $(\preceq 3)$ that $\beta \preceq \alpha$ for all $\alpha \in X$.\\
{\bf ($\preceq$ is inversely well-ranked)} Let $X\not =\emptyset$ and $X\subseteq A$. We will prove by cases:\\
Case 1) There are some $\beta \in X$ such that $\vdash \beta$. Then, by definition of $\preceq$, $\alpha \preceq \beta$ for all $\alpha \in X$.\\
Case 2) All formulae in $X$ are non-tautological. By {\em lower bound} there exists $\beta \in X$ such that $A-\beta \subseteq A-\alpha$ for all $\alpha \in X$. Hence, by definition of $\preceq$, there exists $\beta \in X$ such that $\alpha \preceq \beta$ for all $\alpha \in X$.\qed

{\bf Proof of Theorem \ref{theo_B}}\\
We will prove by cases:\\
Case 1) $\vdash \alpha$. It follows that $A-\alpha=A$ and $(Cn(A)\div_{\leq_{\preceq}}\alpha)\cap A=A$.\\
Case 2) $A\not \vdash \alpha$. It follows that $(Cn(A)\div_{\leq_{\preceq}}\alpha)\cap A=A$ and that $A-\alpha=cut_{\prec}(\alpha)$. By Lemma \ref{lemprev1} (b), it follows that $cut_{\prec}(\alpha)=A$.\\
Case 3)$A\vdash \alpha$ and $\not \vdash \alpha$.\\
We will prove that $A-\alpha=(Cn(A)\div_{\leq_{\preceq}}\alpha)\cap A$ by double inclusion.\\
($\subseteq$) Let $\beta \in A-\alpha$. It follows that $\beta \in A$. It remains to prove that $\beta \in Cn(A)\div_{\leq_{\preceq}}\alpha$, i.e. that $\beta \in \{\psi \in Cn(A): cut_{\preceq}(\psi)\subset cut_{\preceq}(\alpha)\}$.\\
If $\vdash \beta$. It follows trivially by Observation \ref{preTeoA} (b). \\
Assume now that $\not \vdash \beta$. $\beta \in cut_{\prec}(\alpha)$. Hence $cut_{\prec}(\beta) \subset cut_{\prec}(\alpha)$. It follows, from Lemma \ref{obs1} that $cut_{\preceq}(\beta) \subset cut_{\preceq}(\alpha)$.\\
($\supseteq$) Let $\beta\in (Cn(A)\div_{\leq_{\preceq}}\alpha)\cap A$.
If $\vdash \beta$, then it follows from $(\preceq 2)$ that $\{\psi \in A: \beta \preceq \psi\}\subseteq Cn(\emptyset)$. Therefore, since $\not \vdash \alpha$, it follows that $\beta \in cut_{\prec}(\alpha)=A-\alpha$. Assume now that $\not \vdash \beta$. From $\beta\in (Cn(A)\div_{\leq_{\preceq}}\alpha)\cap A$ it follows that $\beta \in A$ and $cut_{\preceq}(\beta) \subset cut_{\preceq}(\alpha)$. Hence there exists $\gamma \in A$ such that $\gamma \in cut_{\preceq}(\alpha)$ and $\gamma \not \in cut_{\preceq}(\beta)$. Hence, $\{\psi\in A:\gamma \prec \psi\}\not \vdash \alpha$ and $\{\psi\in A:\gamma \prec \psi\}\vdash \beta$. Assume by {\em reductio} that $\beta \not \in A-\alpha$ i.e. that $\beta \not \in cut_{\prec}(\alpha)$. Hence, $\{\psi\in A:\beta \preceq \psi\}\vdash \alpha$. From $\{\psi\in A:\beta \preceq \psi\}\vdash \alpha$ and $\{\psi\in A:\gamma \prec \psi\}\not \vdash \alpha$ it follows that $\beta \preceq \gamma$. Therefore, since $\{\psi\in A:\gamma \prec \psi\}\vdash \beta$, it follows that $\{\psi\in A:\beta \prec \psi\}\vdash \beta$ which contradicts $(\preceq 1)$.\qed

{\bf Proof of Theorem \ref{theo_A}}\\
We will prove by cases:\\
Case 1) $\vdash \alpha$. Then $Cn(A)\div_{\leq_{\preceq}}\alpha=Cn(A)$ and $A-\alpha=A$. Hence $Cn(A-\alpha)=Cn(A)=Cn(A)\div_{\leq_{\preceq}}\alpha$. \\
Case 2) $A\not \vdash \alpha$. Then $Cn(A)\div_{\leq_{\preceq}}\alpha=Cn(A)$ and, by Lemma
\ref{lemprev1} (b), $A-\alpha=cut_{\prec}(\alpha)=A$. Hence $Cn(A-\alpha)=Cn(A)=Cn(A)\div_{\leq_{\preceq}}\alpha$. \\
Case 3)$A\vdash \alpha$ and $\not \vdash \alpha$. Hence $Cn(A)\div_{\preceq}\alpha=\{\psi \in Cn(A): \alpha <_{\preceq} \psi\}=\{\psi\in Cn(A): cut_{\preceq}(\psi)\subset cut_{\preceq}(\alpha)\}$.
We will prove that $Cn(A-\alpha)=Cn(A)\div_{\leq_{\preceq}}\alpha$ by double inclusion.\\
($\subseteq$) Let $\beta \in Cn(A-\alpha)$. If $\vdash \beta$, then $\beta \in Cn(A)$ and, by Observation \ref{preTeoA} (b), $cut_{\preceq}(\beta)\subset cut_{\preceq}(\alpha)$. Hence $\beta \in Cn(A)\div_{\preceq}\alpha$.\\
Assume now that $\not \vdash \beta$. From $\beta \in Cn(A-\alpha)$ it follows that $cut_{\prec}(\alpha)\vdash \beta$. Hence, by Lemma \ref{lemprev1} (e), $cut_{\prec}(\alpha\wedge\beta)=cut_{\prec}(\alpha)$. From $\alpha \wedge \beta \vdash \beta$ by Lemma \ref{lemprev1} (c) it follows that $cut_{\prec}(\beta)\subseteq cut_{\prec}(\alpha\wedge \beta)$. Hence $cut_{\prec}(\beta)\subseteq cut_{\prec}(\alpha)$.  From which, together with Lemma \ref{lemprev1} (a) and $cut_{\prec}(\alpha)\vdash \beta$ it follows that  $cut_{\prec}(\beta)\subset cut_{\prec}(\alpha)$. Hence, by Lemma \ref{obs1}, it follows that $cut_{\preceq}(\beta)\subset cut_{\preceq}(\alpha)$. Therefore, since $\beta \in Cn(A)$, it follows that $\beta \in Cn(A)\div_{\leq_{\preceq}}\alpha$. \\
($\supseteq$) Let $\beta \in Cn(A)\div_{\leq_{\preceq}}\alpha$. Hence, $\beta \in Cn(A)$ and $cut_{\preceq}(\beta)\subset cut_{\preceq}(\alpha)$. Assume by {\em reductio} that $\beta \not \in Cn(A-\alpha)$. Therefore $cut_{\prec}(\alpha)\not \vdash \beta$. By Lemma \ref{lemprev1} (f) it follows that $cut_{\prec}(\alpha\wedge \beta)=cut_{\prec}(\beta)$. From $\alpha \wedge \beta \vdash \alpha$, by Lemma \ref{lemprev1} (c), it follows that $cut_{\prec}(\alpha)\subseteq cut_{\prec}(\beta)$. From Observation \ref{preTeoA} (a) it follows that $cut_{\preceq}(\alpha)\subseteq cut_{\preceq}(\beta)$. Contradiction.\qed

{\bf Proof of Theorem \ref{theo_C} }\\
($\Leftarrow$) Let $\div$ be an ensconcement-based withdrawal related to $(A, \preceq)$ and let $\leq=\leq_{\preceq}$. Hence $\div$ satisfies the postulates for severe withdrawals. It remains to show that $\div$ satisfies: {\em base-reduction, upper bound} and {\em lower bound}. \\
{\bf Upper Bound:} Let $\div$ be an ensconcement-based withdrawal related to $(A,\preceq)$. Let $X\not=\emptyset$ and $X\subseteq Cn(A)\setminus Cn(\emptyset)$. From Observation \ref{CorolMAW}, since $(A,\preceq)$ is a bounded ensconcement, it follows that $\leq_{\preceq}$ is bounded. Hence, there exists $\beta \in X$ such that $\beta\leq \alpha$ for all $\alpha \in X$. We will prove that $Cn(A)\div \alpha \subseteq Cn(A)\div \beta$ for all $\alpha \in X$. Let $\gamma \in Cn(A)\div \alpha$. Hence, by definition of $\div$, $\gamma \in Cn(A)$ and $\alpha < \gamma$. By EE1, since $\beta\leq \alpha$ and $\alpha < \gamma$ it follows that $\beta <\gamma$. Hence $\gamma\in Cn(A)\div \beta$. Therefore $Cn(A)\div \alpha\subseteq Cn(A)\div \beta$.\\
{\bf Lower Bound:} Analogous to {\em upper bound}.\\
{\bf Base-reduction:} Let $Cn(A)\div\alpha \vdash \beta$. We will prove that $(Cn(A)\div\alpha) \cap A \vdash \beta$ by cases:\\
Case 1) $\vdash \beta$. Follows trivially. \\
Case 2) $\alpha \not \in Cn(A)$ or $\vdash \alpha$. Follows trivially by ($R_\leq$).\\
Case 3) $\not \vdash \beta, \alpha \in Cn(A)$ and $\not \vdash \alpha$. From $Cn(A)\div\alpha \vdash \beta$ it follows, by ($R_\leq$), that $X\vdash \beta$ where $X=\{\psi \in Cn(A): cut_{\preceq}(\psi) \subset cut_{\preceq}(\alpha) \}$. $X\setminus Cn(\emptyset) \not = \emptyset$, since $\not \vdash \beta$. Let $\psi \in X\setminus Cn(\emptyset)$. Assume that $cut_{\preceq}(\psi)=\emptyset$ and let $\theta \in Cn(\emptyset)$. Hence, by EE5, it follows that $\psi< \theta$. Hence, by Definition \ref{MAWdef2}, $cut_{\preceq}(\theta) \subset cut_{\preceq}(\psi)=\emptyset$. Contradiction. Hence $cut_{\preceq}(\psi)\not=\emptyset$. From Lemma \ref{le1_theo3}, and since $\preceq$ is bounded, it follows that there exists $\delta \in cut_{\preceq}(\psi)$ such that $cut_{\preceq}(\delta)=cut_{\preceq}(\psi)$. Let $Y=\{\mu\in A: cut_{\preceq}(\mu)\subset cut_{\preceq}(\alpha)\}$. Let $\mu_{1}\in Y$ such that $\mu_{1}\preceq \mu$ for all $\mu \in Y$. Let $\lambda \in cut_{\preceq}(\mu_{1})$. Hence $cut_{\preceq}(\lambda) \subseteq cut_{\preceq}(\mu_{1})$, from which follows that $cut_{\preceq}(\lambda) \subset cut_{\preceq}(\alpha)$. Therefore $\lambda \in Y$. Let $\phi\in Y$. It follows that $\mu_{1}\preceq \phi$. Hence $\phi \in cut_{\preceq}(\mu_{1})$. Therefore $Y=cut_{\preceq}(\mu_{1})$. By Lemma \ref{le2_theo3} $cut_{\preceq}(\psi)\vdash \psi$. Hence, since $cut_{\preceq}(\delta)=cut_{\preceq}(\psi)$ it follows that $cut_{\preceq}(\delta) \vdash \psi$. From $cut_{\preceq}(\delta) \subset cut_{\preceq}(\alpha)$ it follows that $\delta \in Y$. Hence $\mu_{1}\preceq \delta$. Therefore $cut_{\preceq}(\delta) \subseteq cut_{\preceq}(\mu_{1})=Y$, and so $Y\vdash \psi$. Hence, for all $\psi \in Cn(A)\div\alpha$ it follows that $Y \vdash \psi$. Therefore, since $Cn(A)\div\alpha \vdash \beta$, it follows that $Y\vdash \beta$. $Y\subseteq (Cn(A)\div\alpha) \cap A$. Hence $(Cn(A)\div\alpha) \cap A \vdash \beta$. \\

($\Rightarrow$) Let $A$ be a belief base and $\div$ be an operator on $Cn(A)$. $\div$ satisfies  $(\div 1)$ to $(\div 4)$, $(\div 6)$, ($\div 9$), {\em base-reduction, upper bound} and {\em lower bound}. Let $\preceq$ be a binary relation on $A$ defined by:\\
 $\alpha \preceq \beta$ if and only if $\alpha \not \in Cn(A)\div \beta$ or $\vdash \beta$. \\
 We will prove that $\preceq$ is a bounded ensconcement.\\
{\bf ($\preceq$1)} Let $\gamma \in A\setminus Cn(\emptyset)$ we must show that $H=\{\alpha \in A: \gamma \prec \alpha\}\not \vdash \gamma$. It is enough to show that $H\setminus Cn(\emptyset)\not \vdash \gamma$. Let $\alpha \in A\setminus Cn(\emptyset)$ and $\gamma \prec \alpha$. Then, $\gamma \preceq \alpha$ and $\alpha \not \preceq \gamma$. Hence, by definition of $\preceq$, it follows that $\gamma \not \in Cn(A)\div \alpha$, $\alpha \in Cn(A)\div \gamma$ and $\not \vdash \gamma$.
$H\subseteq Cn(A)\div \gamma$ where, $\not \vdash \gamma$. Hence, since by $(\div 4)$ $Cn(A)\div \gamma\not \vdash \gamma$ it follows that $H\not \vdash \gamma$.\\
{\bf($\preceq$2)} Let $\alpha,\beta \in A$ be such that $\not \vdash \alpha$ and $\vdash \beta$. From $\vdash \beta$ it follows, by definition of $\preceq$, that $\alpha\preceq \beta$. Assume by {\em reductio} that $\not \vdash \alpha$, $\vdash \beta$ and $\beta \preceq \alpha$. Hence, by definition of $\preceq$, $\beta\not \in Cn(A)\div \alpha$ or $\vdash \alpha$. Contradiction, since $\not \vdash \alpha$ and by ($\div 1$) $\beta \in Cn(A)\div\alpha$. \\
{\bf($\preceq$3)} Follows trivially by definition of $\preceq$.\\
{\bf($\preceq$ is transitive)} Let $\alpha \preceq \beta$ and $\beta \preceq \gamma$. Hence, by definition of $\preceq$, it follows that ($\alpha \not \in Cn(A)\div \beta$ or $\vdash \beta$) and ($\beta \not \in Cn(A)\div \gamma$ or $\vdash \gamma$). Hence, $\alpha \not \in Cn(A)\div \beta$ and ($\beta \not \in Cn(A)\div \gamma$ or $\vdash \gamma$) or ($\vdash \beta$ and ($\beta \not \in Cn(A)\div \gamma$ or $\vdash \gamma$)).
Hence, we have four cases to consider: \\
Case 1) $\alpha \not \in Cn(A)\div \beta$ and $\beta \not \in Cn(A)\div \gamma$. From $(\div 9)$ it follows that $Cn(A)\div \gamma \subseteq Cn(A)\div \beta$. Hence, $\alpha \not \in Cn(A)\div \gamma$. Therefore $\alpha \preceq \gamma$, by definition of $\preceq$.\\
Case 2) $\alpha \not \in Cn(A)\div \beta$ and $\vdash \gamma$. $\alpha \preceq \gamma$ follows trivially by definition of $\preceq$. \\
Case 3) $\vdash \beta$ and $\beta \not \in Cn(A)\div \gamma$. Contradicts $(\div 1)$.\\
Case 4) $\vdash \beta$ and $\vdash \gamma$. $\alpha \preceq \gamma$ follows trivially by definition of $\preceq$.\\
{\bf($\preceq$ is connected)} Let $\alpha \not \preceq \beta$. Hence $\alpha \in Cn(A)\div \beta$ and $\not \vdash \beta$. We will consider two cases:\\
Case 1) $\vdash \alpha$. Hence $\beta \preceq \alpha$, by definition of $\preceq$.\\
Case 2) $\not \vdash \alpha$. Hence, by $\div$ {\em expulsiveness}, $\beta \not \in Cn(A)\div \alpha$. Therefore, by definition of $\preceq$, $\beta \preceq \alpha$.\\
{\bf($\preceq$ is well-ranked)} Let $X\subseteq A$ a non empty set. We will prove by cases:\\
Case 1) $X\subseteq Cn(\emptyset)$. Trivial.\\
Case 2) $X\not \subseteq Cn(\emptyset)$. Let $X'=X\setminus Cn(\emptyset)$. Hence, by $\div$ {\em upper bound} there exists $\beta \in X'$ such that $Cn(A)\div\alpha\subseteq Cn(A)\div\beta$ for all $\alpha \in X'$. By ($\div 4$) $\beta\not \in Cn(A)\div\alpha$ for all $\alpha \in X'$. Hence, by definition of $\preceq$,  there exists $\beta \in X'$ such that $\beta \preceq \alpha$ for all $\alpha \in X'$. If $X=X'$ trivial. Assume now that $X\not=X'$. Let $\gamma \in X\setminus X'$. Hence $\vdash \gamma$ and by $(\preceq 2)$ it follows that $\beta \preceq \gamma$. Therefore, there exists $\beta \in X$ such that $\beta \preceq \alpha$ for all $\alpha \in X$.\\
{\bf($\preceq$ is inversely well-ranked)} Let $X\subseteq A$ a non empty set. We will consider two cases:\\
Case 1)$X\cap Cn(\emptyset)\not=\emptyset$. Let $\beta \in X\cap Cn(\emptyset)$ hence, by definition of $\preceq$, $\alpha \preceq\beta$ for all $\alpha \in X$.\\
Case 2)$X\cap Cn(\emptyset)=\emptyset$. Hence, by $\div$ {\em lower bound}, there exists $\beta\in X$ such that $Cn(A)\div\beta \subseteq Cn(A)\div\alpha$, for all $\alpha \in X$. By ($\div 4$) $\alpha \not \in Cn(A)\div\beta$, for all $\alpha \in X$. Hence, by definition of $\preceq$ there exists $\beta \in X$ such that $\alpha\preceq \beta$, for all $\alpha \in X$.\\
We have proved that $\preceq$ is a bounded ensconcement. Let $\leq_{\preceq}$ be as in Definition \ref{MAWdef2}. According to Observation \ref{Teo4MAW} and Observation \ref{CorolMAW} $\leq_{\preceq}$ is a bounded epistemic entrenchment related to $Cn(A)$. It remains to show that $Cn(A)\div\alpha=Cn(A)\div_{\leq_{\preceq}}\alpha$, where $\div_{\leq_{\preceq}}$ is defined (as in ($R_\leq$))by: \\
$Cn(A) \div_{\leq_{\preceq}} \alpha = \\ \left\{
\begin{array}{ll}
  Cn(A) \cap\{\psi:\alpha<_{\preceq}\psi\}  & $if $ \alpha \in Cn(A)  $and $ \not \vdash \alpha\\
Cn(A)& $otherwise$\\
 \end{array}
\right. $ \\

According to Lemma \ref{obser.19ii} and since $\div$ is a severe withdrawal function, the epistemic entrenchment $\leq$ on which $\div$ is based on is such that: $\alpha \leq \beta$ if and only if $\alpha \not \in Cn(A)\div\beta$ or $\vdash \beta$. Thus to prove that $Cn(A)\div\alpha=Cn(A)\div_{\leq_{\preceq}}\alpha$ it is enough to show that:\\
 $\alpha \leq_{\preceq} \beta$ if and only if $\alpha \not \in Cn(A)\div\beta$ or $\vdash \beta$.\\

{\bf($\Rightarrow$)} Let $\alpha \leq_{\preceq} \beta$. Hence, by definition of $\leq_{\preceq}$,
$\alpha \leq_{\preceq}\beta$ if and only if:\\
i) $\alpha \not \in Cn(A)$, or\\
ii) $\alpha, \beta \in Cn(A)$ and $cut_{\preceq}(\beta)\subseteq cut_{\preceq}(\alpha)$.\\
We will prove by cases:\\
Case 1) $\alpha \not \in Cn(A)$. Then, by $(\div 2)$, $\alpha \not \in Cn(A)\div\beta$. \\
Case 2)  $\alpha, \beta \in Cn(A)$ and $cut_{\preceq}(\beta)\subseteq cut_{\preceq}(\alpha)$.\\
Case 2.1) $\vdash \beta$. Trivial.\\
Case 2.2) $\not \vdash \beta$.\\
$\{\gamma\in A: \{\delta\in A:\gamma \prec \delta\}\not \vdash \beta\} \subseteq \{\gamma\in A: \{\delta\in A:\gamma \prec \delta\}\not \vdash \alpha\}$.\\
Hence, \\
$\{\gamma\in A: \{\delta\in A: (\gamma \not \in Cn(A)\div \delta \textrm{ and } \delta \in Cn(A)\div \gamma \textrm{ and } \not\vdash \gamma) \textrm{ or }(\vdash \delta \textrm{ and } \delta \in Cn(A)\div \gamma \textrm{ and } \not \vdash \gamma)\}\not \vdash \beta\} \subseteq \{\gamma\in A: \{\delta\in A: (\gamma \not \in Cn(A)\div \delta \textrm{ and } \delta \in Cn(A)\div \gamma \textrm{ and } \not\vdash \gamma) \textrm{ or }(\vdash \delta \textrm{ and } \delta \in Cn(A)\div \gamma \textrm{ and } \not \vdash \gamma)\}\not \vdash \alpha\}$.
Therefore according to $(\div 1)$ and $(\div 4)$, \\
$X=\{\gamma\in A: \{\delta\in A: (\gamma \not \in Cn(A)\div \delta \textrm{ and } \delta \in Cn(A)\div \gamma)  \textrm{ or }(\vdash \delta \textrm{ and } \not \vdash \gamma)\}\not \vdash \beta\} \subseteq Y=\{\gamma\in A: \{\delta\in A: (\gamma \not \in Cn(A)\div \delta \textrm{ and } \delta \in Cn(A)\div \gamma) \textrm{ or }(\vdash \delta \textrm{ and } \not \vdash \gamma)\}\not \vdash \alpha\}$.
Assume by {\em reductio} that $\alpha \in Cn(A)\div\beta$. From $\alpha \in Cn(A)\div\beta$ it follows, by {\em base-reduction}, that $Cn(A)\div\beta \cap A \vdash \alpha$. By compactness, there exists a finite subset of $Cn(A)\div\beta \cap A$, $H=\{\alpha_{1},...,\alpha_{n}\}$, such that $H\vdash \alpha$. Let us assume that $H \cap Cn(\emptyset)=\emptyset$. For all $\alpha_{i}\in H$, $\alpha_{i}\in Cn(A)\div\beta=Cn(A)\div\beta'$, for some $\beta' \in A$ (by Observation \ref{AAZ}). Hence, by {\em expulsiveness}, $\beta'\not \in Cn(A)\div\alpha_{i}$. Therefore $\beta'\not\in Y$, since $H\subseteq Z=\{\delta\in A: (\beta' \not \in Cn(A)\div \delta \textrm{ and } \delta \in Cn(A)\div \beta')  \textrm{ or }(\vdash \delta \textrm{ and } \not \vdash \beta')\}$. On the other hand $\beta'\in X$, since $Z\subseteq Cn(A)\div \beta'$, and by ($\div 4$) $Cn(A)\div \beta'\not \vdash \beta$. Hence $X\not\subseteq Y$. Contradiction. \\
{\bf($\Leftarrow$)} Let $\alpha\not\in Cn(A)\div \beta$ or $\vdash \beta$. We will prove by cases:\\
Case 1) $\alpha \not \in Cn(A)$. Trivial.\\
Case 2) $\alpha \in Cn(A)$.\\
Case 2.1) $\vdash \beta$. Then $\alpha, \beta \in Cn(A)$ and $cut_{\preceq}(\beta)\subseteq cut_{\preceq}(\alpha)$.\\
Case 2.2) $\alpha\not\in Cn(A)\div \beta$ and $\not \vdash \beta$. Hence, it follows that $\beta \in Cn(A)$, $\not \vdash \alpha$ and $Cn(A)\div \beta \subseteq Cn(A)\div \alpha$, by ($\div 3$), ($\div 1$) and ($\div 9$), respectively. Let us assume by {\em reductio} that $cut_{\preceq}(\beta)\not\subseteq cut_{\preceq}(\alpha)$. Hence there exists $\psi\in A$ such that $\psi \in cut_{\preceq}(\beta)$ and $\psi \not \in cut_{\preceq}(\alpha)$. From which follows that $\not \vdash \psi$, $C=\{\delta\in A: (\psi \not \in Cn(A)\div \delta \textrm{ and } \delta \in Cn(A)\div \psi)  \textrm{ or }(\vdash \delta \textrm{ and } \not \vdash \psi)\}\not \vdash \beta$ and $C\vdash \alpha$. $C\subseteq Cn(A)\div\psi$. Then $Cn(A)\div\psi \vdash \alpha$. Hence, by ($\div 4$) and {\em linearity}, it follows that $Cn(A)\div\alpha\subset Cn(A)\div\psi$. From $Cn(A)\div \beta \subseteq Cn(A)\div \alpha$ it follows that $Cn(A)\div \beta \subset Cn(A)\div \psi$. By ($\div 9$), $\beta\in Cn(A)\div\psi$. Therefore, by {\em base-reduction}, $Cn(A)\div\psi \cap A\vdash \beta$. On the other hand $Cn(A)\div\psi \cap A \subseteq C$. Hence $C\vdash \beta$. Contradiction.\qed

\bibliographystyle{aaai}

\end{document}